\documentclass[11pt]{article}
\makeatletter
\def\ps@pprintTitle{%
 \let\@oddhead\@empty
 \let\@evenhead\@empty
 \def\@oddfoot{}%
 \let\@evenfoot\@oddfoot}
\makeatother
\usepackage{mathrsfs} 
\usepackage{hyperref,graphicx}
\usepackage[normalem]{ulem}
\usepackage[margin=1.0in]{geometry}
\usepackage{multirow}
\usepackage{booktabs, threeparttable}
\usepackage{subfigure}
\usepackage{physics}
\usepackage{braket}
\usepackage{dsfont}
\usepackage{cite}
\usepackage[cmintegrals,varg]{newtxmath} 

\allowdisplaybreaks[2]
\usepackage{amsfonts}
\usepackage{amsmath,bm}
\usepackage{booktabs, threeparttable}
\usepackage{array}
\newcolumntype{L}[1]{>{\raggedright\arraybackslash}p{#1}}
\usepackage{color}

\usepackage{authblk}

\title{Thermal-response functions and the Peierls-Boltzmann Equation for Second Sound and Phonon Hydrodynamics in Graphene}
\author[1]{Antonio Martinez Margolles}
\author[1,2]{Patrick K. Schelling \thanks{Corresponding author, patrick.schelling@ucf.edu}}
\affil[1]{Department of Physics, University of Central Florida, Orlando, FL 32816-2385, USA}
\affil[2]{Advanced Materials Processing and Analysis Center, University of Central Florida, Orlando, FL 32816-2385, USA}
\begin{document}
\maketitle

\begin{abstract}
We connect
expressions for phonon phase-space distribution functions to microscopic physics of the evolution of heat waves. The role of interference effects that arise as a result of a periodic heating source typically encountered in transient thermal grating (TTG) experiments is then explored. The distribution functions are evaluated as solutions to the Peierls-Boltzmann equation (PBE) in the relaxation-time approximation (RTA). Starting from the PBE, we next develop thermal response functions. The response functions are computed using data from density-functional theory (DFT) calculations. Using this approach, it is shown how solutions to the PBE can be related to the propagation of second phonons as elementary excitations, and within this perspective the necessary conditions for the propagation and observation of second sound is elucidated. The approach developed therefore shows how PBE theory for phonon hydrodynamics and second sound can be modified to properly describe interference and phonon decoherence effects that are likely important at shorter length scales.  Finally, we then discuss how many-body theory can be extended to include coupled scattering channels and hence provide a quantitative theory beyond the RTA.
\end{abstract}

\section{Introduction}

Most efforts to develop theoretical understanding of second sound in solids have been based on solutions of the Peierls Boltzmann equation (PBE), or equivalently the Boltzmann transport equation (BTE)  \cite{Guyer:1966aa,Guyer:1966ab,Hardy:1970aa,Cepellotti_2015,Cepellotti:2016wi,Lee:2017aa,PhysRevB.101.075303,Simoncelli:2020aa,Beardo_2021.2,Sendra:2022aa}. Central to this picture is the use of the phonon distribution function $f_{K}(\bm{r},t)$ which relates to the local density of phonons in the mode $K=(\bm{k},s)$, in which $\bm{k}$ is the phonon wave vector and $s$ labels the phonon band. This theory originated primarily with the works of Guyer and Krumhansl \cite{Guyer:1966aa,Guyer:1966ab} and Hardy \cite{Hardy:1970aa}. The critical assumption made is that normal phonon scattering (N scattering) happens rapidly enough to establish a locally ``displaced'' phonon distribution defined by a drift velocity and a local temperature field. Moreover, umklapp scattering (U scattering) is assumed to be weak such that momentum relaxation is very slow. The resulting ``temperature wave'' is described by a kind of local near equilibrium state, with the corresponding drift velocity and temperature fields propagating according to a damped wave equation. Recently, theory starting from this PBE foundation has continued in the important direction of establishing macroscopic theory based on a set of coupled partial differential equations \cite{Simoncelli:2020aa}.

However, another approach, which dates back to work by Kwok and Martin \cite{Kwok:1966aa}, and somewhat later by Sham \cite{Sham:1967aa,Sham:1967ab}, treats second sound as an elementary excitation. Sham provided a detailed theory based on the interference of two phonon modes which exhibits a pole in their combined Green's function. To highlight the elementary nature of this excitation, Sham  termed these excitations ``second phonons''. Inherent in this picture is that the pole occurs at a frequency $\Omega_{\bm{q}K} = \bm{q} \cdot \bm{v}_{K} $, in which $\bm{v}_{K}$ is the group velocity, and $\bm{q}$ is the wave vector of the second sound excitation, or ``second phonon''. For whatever reason, this alternative theoretical approach has received much less attention than the PBE-based theories.

Recently, we have reported MD simulation results \cite{Bohm:2022aa,MartinezMargolles:2025,Margolles_2026} which were conducted in a regime consistent with existence of coherent oscillations and second phonon propagation as detailed by Sham  \cite{Sham:1967aa,Sham:1967ab}. To observe this phenomena, the necessary requirement is that the phonon lifetime is long enough such that multiple periods of second phonon oscillation can occur without losing phase coherence. In graphene and hBN \cite{MartinezMargolles:2025,Margolles_2026}, we have demonstrated these conditions are met for length scales $\sim 100$nm and temperatures $T=300$K and lower. An important result of these simulations was the realization that second sound propagates with a bandwidth in part determined by the range frequencies $\Omega_{\bm{q}K} = \bm{q} \cdot \bm{v}_{K} $. As the magnitude $|\bm{q}|$ increases, the bandwidth also increases, potentially becoming the primary factor in the lifetime of second-sound oscillations. 

In this paper, we make a connection between these two perspectives. Specifically, we examine from a microscopic perspective how one might define the  phonon distribution function $f_{K}(\bm{r},t)$. We show how this necessarily involves phonon interference, and hence becomes directly connected to the ``second phonons'' described by Sham \cite{Sham:1967aa,Sham:1967ab}. The derived distribution functions propagate as damped waves and are solutions to the PBE for each mode in the relaxation-time approximation (RTA). Observation of second sound requires phase coherence which is limited by scattering and the bandwidth of the second phonons. The properties of the phonons, both their linewidths and group velocities, are computed using density-functional theory (DFT) calculations. The bandwidth of second sound is computed using linear-response theory and thermal response functions. In this way, we have established a simple approach to predict conditions where second sound can be observed in a transient thermal grating (TTG) experiment. Additional connections between this approach and the Green's functions at the lowest level of approximation are established, with an explanation of how vertex corrections should result in a more complete picture. Moreover, we identify conditions where the propagation of the so-called second phonons as elementary excitations do not provide a reliable picture. Finally, we establish that previous PBE-based theories are likely to fail at shorter length scales, while remaining predictive of phonon hydrodynamics and second sound at longer length scales.

\section{Background and motivation}

Second sound is a feature of transport in a hydrodynamic regime.
There, it is presumed that N scattering occurs at a significantly higher rate than U scattering, resulting in a displaced
phonon distribution. This condition is certainly satisfied in materials at low temperatures, and appears to be especially relevant in two-dimensional materials including graphene and monolayer hBN. Given sufficiently rapid N scattering, the phonons are expected to attain the so-called ``displaced'' or  ``drifting'' distribution.  This distribution is characterized by a vector ``drift velocity'' field $\bm{V}(\bm{r},t)$ and scalar temperature
field $T(\bm{r},t)$, with the resulting expression,
\begin{equation} \label{displaced}
N^{(disp)}_{\bm{k} s}(\bm{r},t) = \left[\exp{\frac{\hbar \left( \omega_{\bm{k}s} - \bm{V}(\bm{r},t) \cdot \bm{k} \right)} {k_{B}T(\bm{r},t)}} - 1 \right]^{-1} \text{  ,}
\end{equation}
Assuming the perturbing fields are small, the distribution function can be expanded,
\begin{equation} \label{displaced2}
N^{(disp)}_{K}(\bm{r},t) \approx N_{K} +{ \hbar \omega_{K} \over k_{B}T_{0}}  N_{K}(N_{K}+1) \left[{\bm{V}(\bm{r},t) \cdot \bm{k} \over \omega_{K} }+ {\delta T(\bm{r},t) \over T_{0}}\right]
\text{  ,}
\end{equation}
in which $N_{K}$ is the Bose-Einstein distribution function at temperature $T_{0}$. Notably, the two terms in the expansion are related to eigenstates of the N scattering operator with zero eigenvalue.

In second sound theories starting from this perspective, the above distribution is coupled with PBE theory in various approximations, and, along with the energy and momentum-balance equations, damped wave equations for $\bm{V}(\bm{r},t)$ and $T(\bm{r},t)$ are obtained \cite{Guyer:1966aa,Guyer:1966ab,Hardy:1970aa,Lee:2017aa,Simoncelli:2020aa}. Moreover, use is made of the fact that expansion of Eq. \ref{displaced} results in four eigenvectors of the N scattering matrix \cite{Guyer:1966aa,Guyer:1966ab}.

Central to these previous works is the assumption that the fields $\bm{V}(\bm{r},t)$ and $T(\bm{r},t)$ propagate as a wave with a single frequency and lifetime. The finite lifetime for second sound is then determined entirely by the anharmonic phonon scattering rates. In the following, we suggest that the propagation of a temperature and drift-velocity field that is distinct for each phonon mode $K$. The frequency of each of these fields is given for each mode $K$ by the relation $\Omega_{\bm{q}K} = \bm{q}\cdot\bm{v}_{K}$, in agreement with the ``second phonon'' concept of Sham \cite{Sham:1967aa,Sham:1967ab} and previous MD simulations from our group \cite{Bohm:2022aa,MartinezMargolles:2025,Margolles_2026}. Because these fields propagate with a range of frequencies $\Omega_{\bm{q}K}$, they tend to lose phase coherence over time and hence limit the lifetime of second sound. This is expected to be important especially for TTG experiments with shorter spatial periodiciy as might be expected from recent experiments using UV lasers.

\section{Peierls-Boltzmann Equation for Second Sound}

In the following, we focus on studying heat waves that occur due to a source that does not generate any net quasi-momenta. This is expected, for example, in a TTG experiment. Both N and U scattering
play a role in establishing equilibrium. We provide a connection between the phonon distribution function and the microscopic phonon physics.  The distribution function for each mode $K=(\bm{k},s)$ is obtained as solution to the PBE and propagates as damped wave. The oscillations are thus characterized by a local amplitude (much like a local temperature field), and a vector displacement field. However, in contrast to previous works, these fields evolve independently for each phonon mode.

The PBE in the relaxation-time approximation (RTA) is given by,
 \begin{equation} \label{PBE1}
 {\partial f_{K} \over \partial t} + \bm{v}_{K} \cdot \bm{\nabla} f_{K} = -{f_{K} - f_{K}^{(0)} \over \tau_{C}}
 \text{  ,}
 \end{equation}
 in which $K=(\bm{k}s)$ represents a phonon mode with wave vector $\bm{k}$ in branch $s$, $\bm{v}_{K}$ is the group velocity of the mode, and $f_{K}^{(0)}$ is equilibrium phonon distribution given by the Bose-Einstein distribution function
 defined at temperature $T_{0}$. The scattering rate is the sum of normal and umklapp scattering rates,
 \begin{equation}
 {1 \over \tau_{C}} =  {1 \over \tau_{N}}+ {1 \over \tau_{U}}  \text{   ,}
 \end{equation}
 in which the value of the scattering rate depends on the particular mode $K$.
 
 The quantity $f_{K}(\bm{r},t)$, the phonon distribution function,  is used to represent the local phonon distribution for mode $K$ at point $\bm{r}$ and time $t$. We define another quantity $g_{K}(\bm{r},t) = f_{K}(\bm{r},t) - f_{K}^{(0)}$ which describes the deviations from equilibrium. Since $f_{K}^{(0)}$ is independent of time and spatial coordinate, the PBE becomes,
 \begin{equation} \label{PBE2}
 {\partial g_{K} \over \partial t} + \bm{v}_{K} \cdot \bm{\nabla}g_{K} = -{g_{K} \over \tau_{C}}
 \text{  ,}
 \end{equation}

The  connection  between $g_{K}(\bm{r},t)$ and microscopic physics is not uniquely defined. Some insight is obtained by realizing that the local deviation function  $g_{K}(\bm{r},t)$ depends  on phonon interference effects. Strictly speaking, phonons are localized in $\bm{k}$ space and completely delocalized in position space. To generate a spatially varying deviation function  $g_{K}(\bm{r},t)$, we require some delocalization in $\bm{k}$ space. In Appendix A, a starting point based on the harmonic hamiltonian is used to develop a reasonable approach to define $g_{K}(\bm{r},t)$ corresponding to a heat wave with wave vector $\bm{q}$. Including scattering within the RTA, we show in Appendix A the we can take,
  \begin{equation} \label{PBEsoln}
g_{K}(\bm{r},t) = N_{K} \cos{\left(\bm{q} \cdot \bm{r}- \Omega_{\bm{q}K}t \right)} e^{-{t \over \tau_{C}}}
 \end{equation}
 Substitution into Eq. \ref{PBE2} shows that it is an exact solution of the PBE. The frequency $\Omega_{\bm{q}K} = \bm{q} \cdot \bm{v}_{K} \approx\omega_{\bm{k}+ {1 \over 2} \bm{q},s} -\omega_{-\bm{k}+ {1 \over 2} \bm{q},s} $ represents the ``beating'' between two interfering modes in phonon branch $s$, and $N_{K}$ represents the number of phonons in the state $K$. Finally, the picture of two interfering modes is exactly the ``second phonon'' elementary excitation identified by Sham \cite{Sham:1967aa,Sham:1967ab}, and the resonant frequency $\Omega_{\bm{q}K} $ is the locations of the pole in its Green's function. This point was also elaborated in the theory developed in Ref. \cite{Schelling:2025aa}.
 
 We next assume that $N_{\bar{K}}=N_{K}$, with $\bar{K} = (-\bm{k}s)$, which is consistent with the assumption of zero net quasi-momenta. This then results in a standing wave,
   \begin{equation} \label{PBEsoln2}
g_{K}(\bm{r},t) +  g_{\bar{K}}(\bm{r},t)  = 
2N_{K} \cos { \left( \bm{q} \cdot \bm{r} \right) } 
 \cos{\left( \Omega_{\bm{q}K}t \right)} e^{-{t \over \tau_{C}}}
 \end{equation}
 Thus one expects standing waves with local regions of high and low energy density. This is much like what occurs with the temperature field $T(\bm{r},t)$, however the oscillation frequency is specific to each mode $K$. To establish the local displacement field which captures the local quasi-momentum deviation, we take the difference,
    \begin{equation} 
 g_{K}(\bm{r},t) -  g_{\bar{K}}(\bm{r},t)  = 
 2N_{K} \sin { \left( \bm{q} \cdot \bm{r} \right) } 
 \sin{\left( \Omega_{\bm{q}K}t \right)} e^{-{t \over \tau_{C}}}
  \end{equation}
 and relate to the expansion in Eq. \ref{displaced2}, we obtain the local displacement field,
  \begin{equation} 
 \bm{V}_{\bm{q}K}(\bm{r},t) \cdot {\bm{k} \over k}  = 
 {k_{B}T_{0} \over \hbar k (N_{K}+1)} 
 \sin { \left( \bm{q} \cdot \bm{r} \right) } 
 \sin{\left( \Omega_{\bm{q}K}t \right)} e^{-{t \over \tau_{C}}}
   \end{equation}
   Importantly, for a given heat wave excitation wave vector $\bm{q}$, this vector field is a standing wave that evolves distinctly for each
   phonon state $K$. This is in contrast to the standard PBE theory of second sound where $\bm{V}(\bm{r},t)$ is the only local displacement field which oscillates with a single frequency.

The contribution to a heat-current density resulting from an excitation with wave vector $\bm{q}$ due to phonon mode $K$ is,
\begin{equation} \label{current}
\bm{J}(\bm{r},t) = {1 \over V}  
N_{K} \hbar \omega_{K} \bm{v}_{K} 
\left[ \cos{\left(\bm{q} \cdot \bm{r}- \Omega_{\bm{q}K}t +\Delta \phi_{\bm{q}K} \right)} e^{-2 \Gamma_{K}t}\right]
\text{   ,}
\end{equation}
 in which the scattering rate ${1 \over \tau_{C}}$ is written in terms of the phonon linewidth  $ \Gamma_{K}$. For generality, we have added a phase factor $\Delta \phi_{\bm{q}K} $. Without an external excitation, these phase factors for different modes $K$ are random, and hence, after summation over the entire spectrum of modes $K$, collective second-sound currents will not be observed. In the presence of an external source, for example interfering lasers in a TTG experiment, phase coherence is generated across the spectrum, and the phase factors are no longer random. Anharmonic scattering rates dissipate second sound. However, another factor, and a key insight of this paper, is that phase coherence is lost over time due to the wide range of frequencies $\Omega_{\bm{q}K}$ present in a heat wave. 
 
 \section{Thermal response functions}
 
 Here we derive thermal response functions starting from the solutions to the PBE with the objective of computing the spectrum of second-sound oscillations \cite{Fernando_2020,Bohm:2022aa,MartinezMargolles:2025,Margolles_2026}. The analysis is done in Fourier space which is consistent with excitations by a spatially-periodic heat source. When a heat perturbation $U_{\bm{q}}$ introduced at an initial time, a heat-current density $J_{\bm{q}}(t)$ results at a later time $t$. This is given by the linear-response expression,
  \begin{equation} \label{response}
  J_{\bm{q}}(t) = -{i \over c_{V}} q K_{\bm{q}}(t) U_{\bm{q}}(0)
  \end{equation}
  in which $\bm{q}$ is chosen along a principal axis of the crystal and hence $ J_{\bm{q}}(t)  = \bm{J}_{\bm{q}}(t) \cdot {\bm{q} \over q}$. It is assumed next that $N_{K}=N_{\bar{K}}$ and then the perturbation is determined by the deviations from equilibrium $g_{K}(\bm{r},t)$. The objective here is to develop a PBE theory for the response function $K_{\bm{q}}(t)$.
  
  To explicitly describe a situation which corresponds to a TTG experiment, we assume an initial sinusoidal energy-density profile. We can then represent the energy density at Bravais lattice vector $\bm{R}_{l}$ with,
  \begin{equation}
      U_{l}(t=0) = U_{\bm{q}}(0)e^{i \bm{q} \cdot \bm{R}_{l}}+U_{-\bm{q}}(0)e^{-i \bm{q} \cdot \bm{R}_{l}}
      \text{   .}
  \end{equation}
 We next assume,
  \begin{equation}
  U_{\bm{q}}(0) =U_{-\bm{q}}(0)= {1 \over 2V} \sum_{K} N_{K} \hbar \omega_{K}
  \end{equation}
  and then with the assumption $N_{K}=N_{\bar{K}}$, the  energy density in real space at $t=0$ is (see Appendix A),
    \begin{equation}
      U_{l}(t=0)-U_{0} ={1 \over 2V} \sum_{K} N_{K} \hbar \omega_{K} \cos{\left(\bm{q} \cdot \bm{R}_{l} \right)} 
    \text{   ,}   \end{equation}
  in which $U_{0}$ represents the $\bm{q}=0$, uniform energy density. Hence, the perturbation generates a standing wave with zero net quasi-momentum. Now, given this initial perturbation, the heat-current density evolves according to (see Appendix A),
  \begin{equation}
  J_{l}(t) = J_{\bm{q}}(t)e^{i\bm{q}\cdot \bm{R}_{l}}+J_{-\bm{q}}(t)e^{-i\bm{q}\cdot \bm{R}_{l}}
  =\frac{1}{V}\sum_{K}N_{K}\hbar \omega_{K} v_{K}\sin{\left(\bm{q} \cdot \bm{R}_{l} \right)}
  \sin{\left(\Omega_{\bm{q}K}t \right)}e^{-2 \Gamma_{K}t}
      \end{equation}
  Then the heat-current Fourier components are,
    \begin{equation}
    J_{\bm{q}}(t) = -J_{-\bm{q}}(t) = -{i \over 2V} \sum_{K} N_{K} \hbar \omega_{K} v_{K} \sin{\left(\Omega_{\bm{q}K}t \right)}e^{-2 \Gamma_{K}t}
    \end{equation}
    in which $v_{K} = \bm{v}_{K} \cdot {\bm{q} \over q}$ is the component of the group velocity $\bm{v}_{K}$ of mode $K$ along the perturbation vector $\bm{q}$. From this, comparison with the definition in Eq. \ref{response} yields for the response function for $t \geq 0$,
      \begin{equation}  
    K_{\bm{q}}(t) = \left({c_{V} \over q} \right) {\sum_{K} N_{K} \hbar \omega_{K} v_{K} \sin{\left(\Omega_{\bm{q}K}t \right)}e^{-2 \Gamma_{K}t}  \over \sum_{K} N_{K} \hbar \omega_{K}}
     \end{equation}
     with $K_{\bm{q}}(t)=0$ for $t < 0$.
   This expression can be Fourier transformed to obtain the response function $\tilde{K}_{\bm{q}}(\omega) = K^{\prime}_{\bm{q}}(\omega) + iK^{\prime \prime}_{\bm{q}}(\omega)$ in terms of its real and imaginary parts. Computing these,
       \begin{equation}    \label{kprime}
  K^{\prime}_{\bm{q}}(\omega) =    
  \left({c_{V} \over  2q\sum_{K} N_{K} \hbar \omega_{K} } \right) \sum_{K} N_{K} \hbar \omega_{K} v_{K}
  \left[ {\omega + \Omega_{\bm{q}K} \over (\omega+\Omega_{\bm{q}K})^{2} + 4\Gamma_{K}^{2}} -
   {\omega - \Omega_{\bm{q}K} \over (\omega-\Omega_{\bm{q}K})^{2} + 4\Gamma_{K}^{2}}
  \right] 
          \end{equation}
       \begin{equation}    \label{kprimeprime}
  K^{\prime \prime}_{\bm{q}}(\omega) =    
  \left({c_{V} \over  2q\sum_{K} N_{K} \hbar \omega_{K} } \right) \sum_{K} N_{K} \hbar \omega_{K} v_{K}
  \left[ - { 2 \Gamma_{K} \over (\omega+\Omega_{\bm{q}K})^{2} + 4\Gamma_{K}^{2}} +
   {2 \Gamma_{K} \over (\omega-\Omega_{\bm{q}K})^{2} + 4\Gamma_{K}^{2}}
  \right] 
          \end{equation}
          These expressions make clear the presence of poles at $\omega = \pm \Omega_{\bm{q}K}$ in the second-sound spectrum. They
          also follow the expected symmetries, $  K^{\prime}_{\bm{q}}(\omega)  =   K^{\prime }_{\bm{q}}(-\omega) $ and   $K^{\prime \prime}_{\bm{q}}(\omega) = -  K^{\prime \prime}_{\bm{q}}(-\omega)$. 

          \begin{figure}
\begin{centering}
\includegraphics[scale=0.75]{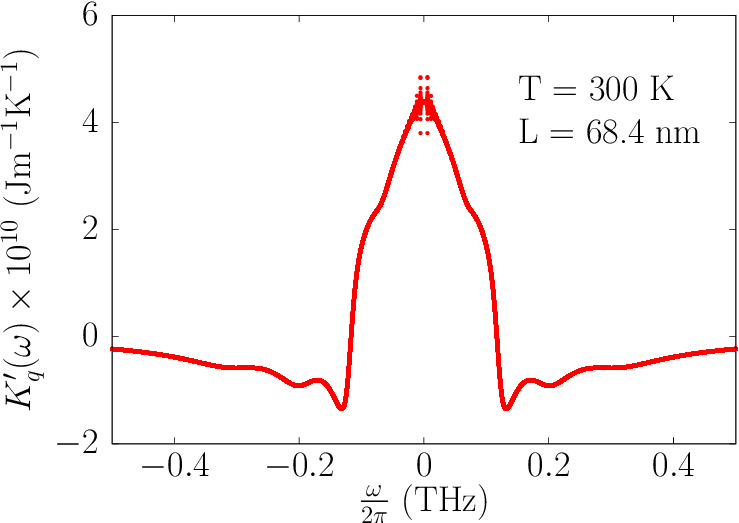} 
\caption{Real part $K_{\bm{q}}^{\prime}(\omega)$ of the response function as a function of frequency ${\omega \over 2\pi}$ at $T=300K$. Calculation
for excitation vector $\bm{q}$, with $L={2 \pi \over q}=68.4$nm.
}
\label{300K_real}
\end{centering}
\end{figure}
     
     \begin{figure}
\begin{centering}
\includegraphics[scale=0.75]{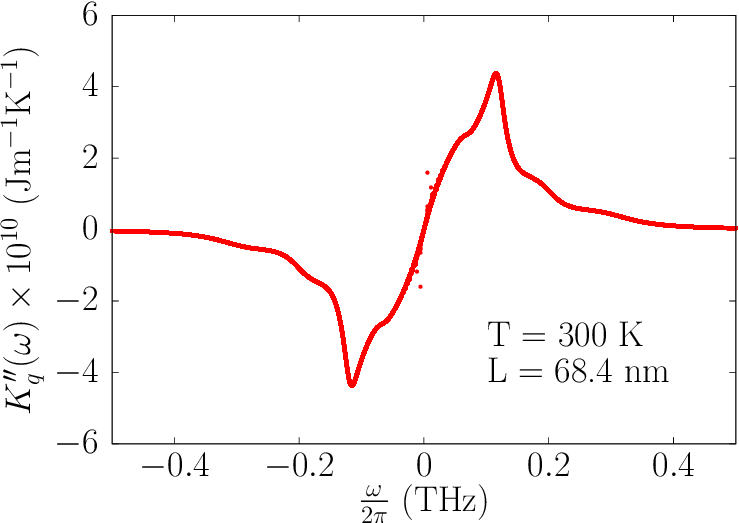}  
\caption{Imaginary part $K_{\bm{q}}^{\prime \prime}(\omega)$ of the response function as a function of frequency ${\omega \over 2\pi}$ at $T=300K$. Calculation
for excitation vector $\bm{q}$, with $L={2 \pi \over q}=68.4$nm.
}
\label{300K_imag}
\end{centering}
\end{figure}
     
     \begin{figure}
\begin{centering}
\includegraphics[scale=0.75]{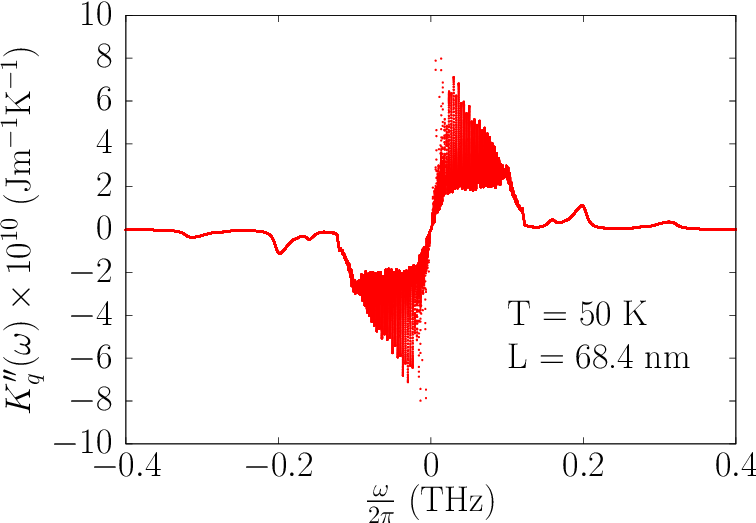} 
\caption{Imaginary component $K_{\bm{q}}^{\prime \prime}(\omega)$ determined at $T=50$K. 
In contrast to the $T=300$K results, sharp peaks are evident in the spectra, potentially consistent with ``ballistic'' transport. Calculation
for excitation vector $\bm{q}$, with $L={2 \pi \over q}=68.4$nm.
}
\label{50K_imag}
\end{centering}
\end{figure}

          
We have computed $\tilde{K}_{\bm{q}}(\omega)$ using Eqs. \ref{kprime}-\ref{kprimeprime} with DFT data provided by the VASP code \cite{Kresse_1994,KRESSE199615,Kresse:1999aa,PhysRevB.54.11169}. The system was graphene with the lattice parameter fixed at the $T=0$K value $a_{0}= 1.425 \AA$. Electronic-structure calculations were performed for a $8 \times 8 \times 1$ supercell with 128 atoms. In reciprocal space, a $3 \times 3 \times 1$ Monkhorst-Pack mesh was used with Gaussian smearing for Brillouin-zone integration. The phono3py code was then used to determine the phonon spectra, group velocities, and scattering rates on a $120 \times 120 \times 1$ grid in reciprocal space using the tetrahedron method for Brilloun-zone integration \cite{Togo_2023.2,Togo_2023}.  Using the resulting data, we computed $\tilde{K}_{\bm{q}}(\omega)$ for different vectors $\bm{q}$ chosen in the $\Gamma$-M direction.

In Figs. \ref{300K_real}-\ref{300K_imag}, the real and imaginary parts of the response function are shown for $T=300K$ and $q = 0.0919$ rad nm$^{-1}$ corresponding to the periodicity length $L = 68.4$ nm. The results look very similar to our recent MD simulation results reported in Ref. \cite{Margolles_2026} which were obtained at the same temperature and excitation vector $\bm{q}$. In graphene, the largest frequency $\Omega_{\bm{q}K}$ occurs for long wavelength LA modes, leading to poles at $\pm 0.31$THz. This frequency basically defines the limits (i.e. the bandwidth) of the spectral response function. The somewhat wider spectrum reflects peak broadening due to anharmonic scattering. Finally, while the structure of the response function does not involve sharp peaks, it nevertheless demonstrates clear structure. This structure reflects the poles at $\omega=\Omega_{\bm{q}K}$ in the spectra.

Using the same vector $\bm{q}$, the results in Fig. \ref{50K_imag} for $K_{\bm{q}}^{\prime \prime}(\omega)$ at $T=50K$ show essentially the same overall spectral width. However, in contrast to $T=300$K, sharper, more distinct features appear in the spectrum, with noticeable peaks due to longer phonon lifetimes. In fact, the very sharp peaks evident in the spectra are due to the finite spacing of the $120 \times 120 \times 1$ reciprocal-space grid and hence represent a kind of ``finite-size'' effect. With a finer grid, the peaks would smear together, although clear structure in the response function would nevertheless still be evident. It might be said that rather than representative of second sound in a hydrodynamic regime, Fig. \ref{50K_imag} exhibits aspects of ``ballistic'' transport. However, regardless of how these concepts are defined, we expect, and will shortly demonstrate, that a wave-like heat current will result in these conditions.

\begin{figure}
\begin{centering}
\includegraphics[scale=0.75]{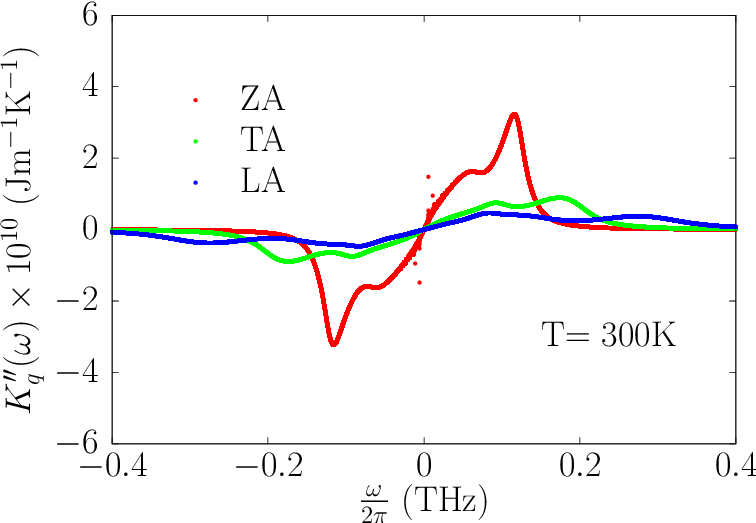} 
\caption{Contributions of the acoustic modes to $K_{\bm{q}}^{\prime \prime}(\omega)$ determined at $T=300$K for excitation vector $\bm{q}$, with $L=\frac{2\pi}{q}=68.4$nm.}
\label{300K_imag_bands}
\end{centering}
\end{figure}

\begin{figure}
\begin{centering}
\includegraphics[scale=0.75]{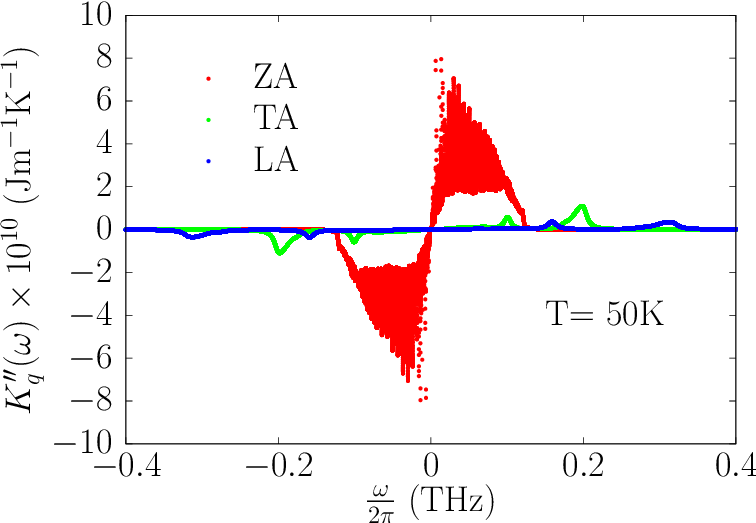} 
\caption{Contributions of the acoustic modes to $K_{\bm{q}}^{\prime \prime}(\omega)$ determined at $T=50$K. Calculation done for excitation vector $\bm{q}$, where $L=\frac{2\pi}{q}=68.4$nm. }
\label{50K_imag_bands}
\end{centering}
\end{figure}

The contributions due to different phonon branches were determined at $T=300$K. These are shown in Fig. \ref{300K_imag_bands}. Clearly ZA modes make the dominant contribution, although resolved peaks for TA and LA modes are also evident. Summation over each of the branch contributions results in the rather broad, somewhat featureless spectra in Fig. \ref{300K_imag}. At $T=50$K, unsurprisingly the ZA modes are still dominant, while TA and LA peaks are more sharply resolved. This is shown in Fig. \ref{50K_imag_bands}. Hence, at both $T=50$K and $T=300$K, each phonon branch makes a distinct contribution to the spectra, yet with ZA phonons dominating transport. 

\begin{figure}
\begin{centering}
\includegraphics[scale=0.75]{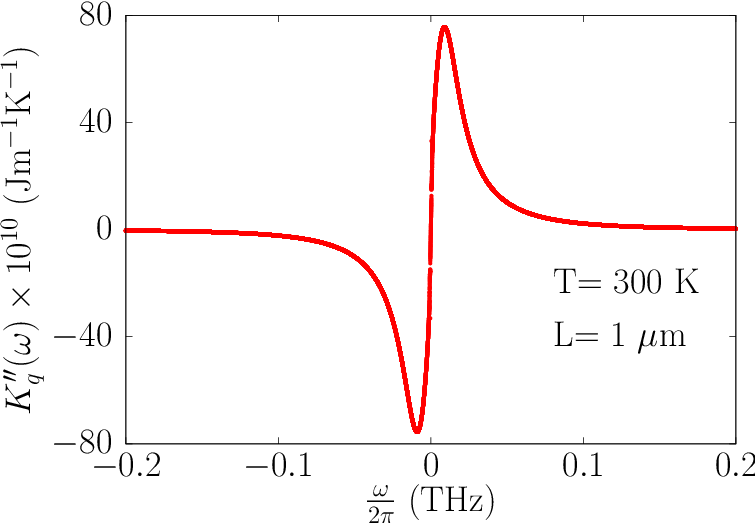} 
\caption{Imaginary part $K_{\bm{q}}^{\prime \prime}(\omega)$ of the response function as a function of frequency ${\omega \over 2\pi}$ for system length $L=1 \mu m$ at  $T=300$K. 
}
\label{300K_imag_1micron}
\end{centering}
\end{figure}

\begin{figure}
\begin{centering}
\includegraphics[scale=0.75]{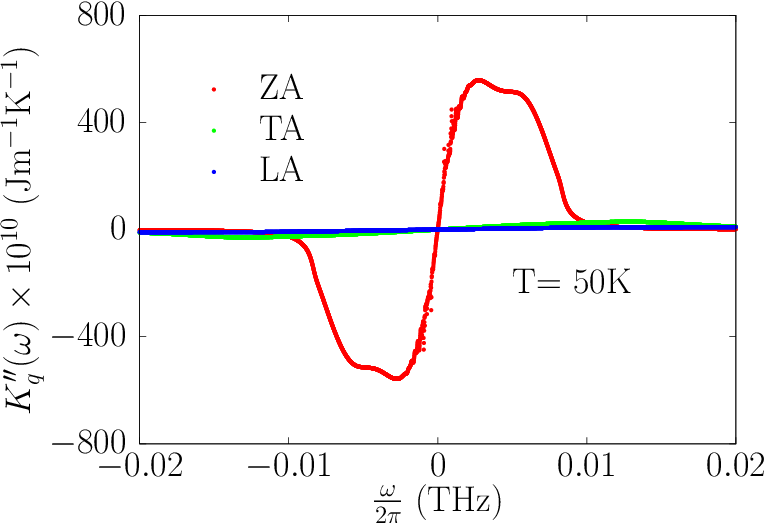}
\caption{Contributions of the acoustic modes to $K_{\bm{q}}^{\prime \prime}(\omega)$ determined at $T=50$K for system length $L=1\mu$m. }
\label{50K_imag_bands_1um}
\end{centering}
\end{figure}

Next, using the same data computed on the $120 \times 120 \times 1$ grid in reciprocal space, the response functions were computed at a smaller value for $\bm{q}$. In particular, we use a value $L=\frac{2\pi}{q}=1\mu$m to predict behavior at a much longer scale consistent with previously-reported TTG experiments. Since the poles occur at $\Omega_{\bm{q}K}=\bm{q} \cdot \bm{v}_{K}$, the spectral width of the response is narrower is comparison to the $L=68.4$nm results. In Fig. \ref{300K_imag_1micron} for T=300K, we show only the imaginary component $K_{\bm{q}}^{\prime \prime}(\omega)$. At $T=300$K, the spectra is very broad and featureless, demonstrating a dominant role for phonon scattering in determining the spectral width of the response. In this case, second sound is not evident as is demonstrated and discussed later. For the same
$\bm{q}$, the results for $T=50$K are shown in Fig. \ref{50K_imag_bands_1um}. The results in Fig. \ref{50K_imag_bands_1um} are broken down by contributions of the different acoustic bands to demonstrate the dominant role of ZA phonons. At $L=1\mu$m and $T=50$K, the spectral response of LA and TA modes shows a lack of coherent second phonon propagation.  However, second phonon propagation in the ZA branch is evident in this case, as we demonstrate later by showing the time-dependent response. 

To summarize the above observations, for $L=1\mu$m and $T=50$K, the picture of second phonon propagation appears to be relevant only in the ZA phonon branch. For the ZA phonons, the spectral width depends mostly on the range of frequencies $\Omega_{\bm{q}K}$. For the same scale but at $T=300$K, coherent second phonons do not propagate as elementary excitations, and the spectral width is primarily controlled by the scattering rate $\Gamma_{K}$ for the phonons.

\begin{figure}
\begin{centering}
\includegraphics[scale=0.75]{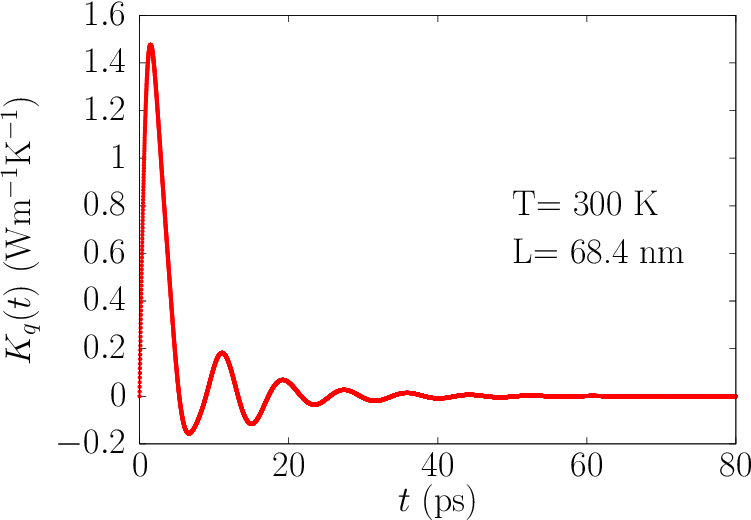} 
\caption{$K_{\bm{q}}(t)$, the time dependent response function, evaluated for $T=300$K and $L=68.4$nm}
\label{temporal_300K_waves}
\end{centering}
\end{figure}

\begin{figure}
\begin{centering}
\includegraphics[scale=0.75]{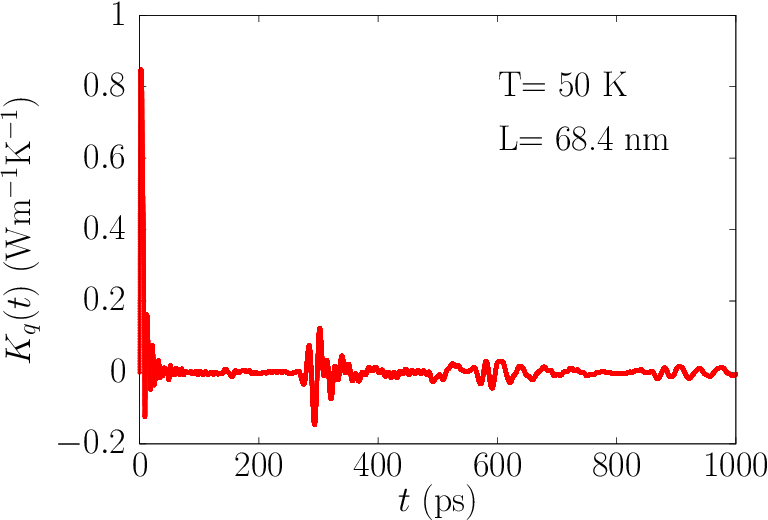} 
\caption{$K_{\bm{q}}(t)$ for $T=50$K and $L=68.4$nm}
\label{temporal_50K}
\end{centering}
\end{figure}

The propagation of second sound waves can be established by plotting the response function $K_{\bm{q}}(\tau)$. In Fig. \ref{temporal_300K_waves}, $K_{\bm{q}}(\tau)$ is plotted for $T=300$K and $L=68.4$nm. The results
are quite close to the MD simulation results in Ref. \cite{Margolles_2026}, demonstrating oscillations with period $\sim 8-9$ps and rather strong damping. Based on the analysis above, the heat waves are primarily carried by ZA modes, and the strong damping is explained by the rather wide range of frequencies $\Omega_{\bm{q}K}$ present for the ZA branch. For the same length $L=68.4$nm at the lower temperature $T=50$K, oscillations persist beyond $1 \mu$s and are much more complicated, reflecting the richer spectra. This is shown in Fig. \ref{temporal_50K}. Interestingly, while damping still occurs quite rapidly at $T=50$K due to phonon decoherence, at later times phase coherence can recur. For example, near $300$ps in Fig. \ref{temporal_50K}, the amplitude of the oscillations briefly increases.

\begin{figure}
\begin{centering}
\includegraphics[scale=0.75]{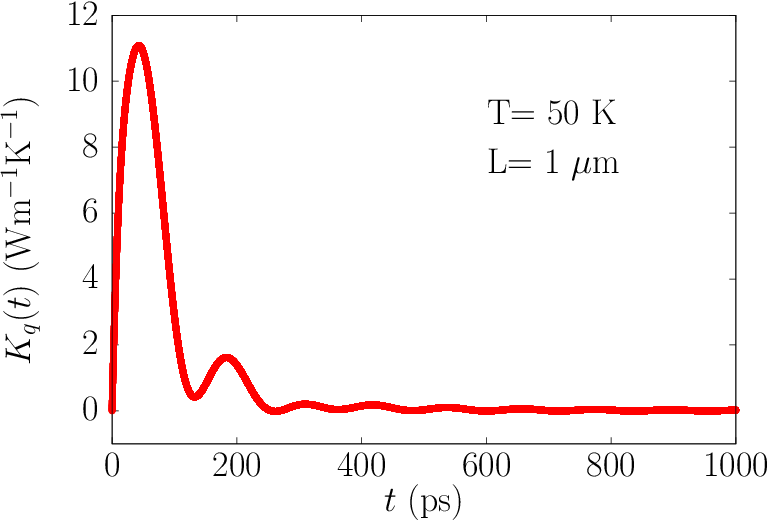} 
\caption{$K_{\bm{q}}(t)$ for $T=50$K and $L=1\mu m$}
\label{temporal_50K_1micron}
\end{centering}
\end{figure}

At the scale $L=1\mu$m, the second phonon frequencies $\Omega_{\bm{q}K}$ are much lower and are more likely to be comparable to the scattering rates $\Gamma_K$. Hence, for smaller $\bm{q}$ (larger $L$) coherent wave propagation and second sound may not be observed. For $T=50$K and $L=1\mu$m, Fig \ref{temporal_50K_1micron} still exhibits oscillations, but with substantially stronger damping. For $T=300$K and $L=1\mu$m, $K_{\bm{q}}(\tau)$ exhibits rapid decay without oscillations, consistent with the very broad spectral response. We do not show this result here in a figure.

  \section{Second phonon as an elementary excitation}
  Here we examine the conditions for second phonon propagation with the perspective that it is a minimal requirement for the observation of  second sound.
  The requirement for an elementary excitation to be a useful concept is that the lifetime should be significantly greater than the period of oscillation.  Here we examine the ratio $R_{\bm{q}K}$ given by,
  \begin{equation}
  R_{\bm{q}K}={4 \pi \Gamma_{K} \over |\Omega_{\bm{q} K}|}
  \end{equation}
  in which the absolute value bars in the denominator are added since the frequencies $\Omega_{\bm{q} K}$ can take negative values.
  For modes $K$ where $R_{\bm{q}K}>1$, scattering is strong enough so that transport in mode $K$ does not follow the paradigm of an elementary excitation. By contrast, for values $R_{\bm{q}K} \lesssim 1$, second-phonon oscillations should be observable for mode $K$. This is a minimal requirement for observing second sound, since even when many modes satisfy $R_{\bm{q}K} \lesssim 1$, second sound can involve propagation over a broad range of frequencies $\Omega_{\bm{q}K}$ with the tendency to lose phase coherence with time.

\begin{figure}
\begin{centering}
\includegraphics[scale=0.75]{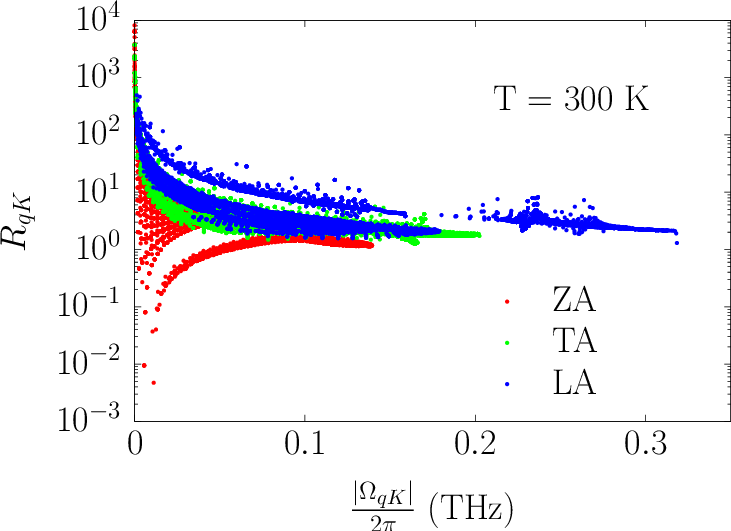} 
\caption{Ratio $R={4 \pi \Gamma_{K} \over |\Omega_{\bm{q} K}|}$ calculated at T = 300K for the three acoustic phonon bands. The length scale of the system is $L=68.4$nm.}
\label{R_Omegaqks_300K}
\end{centering}
\end{figure}

\begin{figure}
\begin{centering}
\includegraphics[scale=0.75]{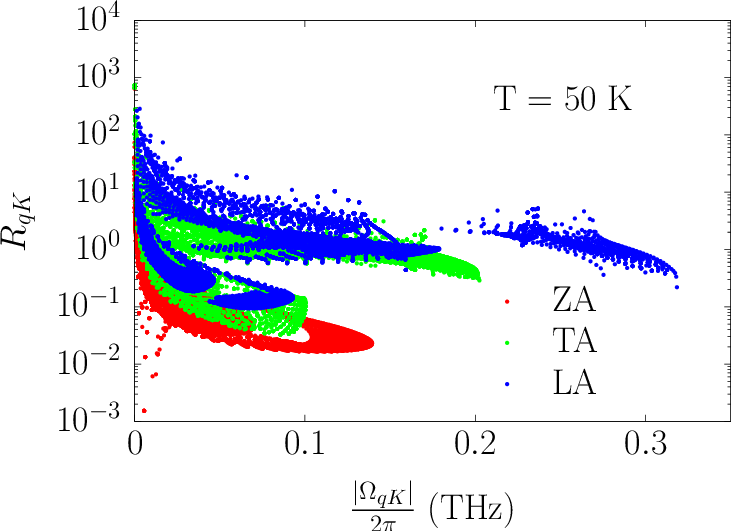} 
\caption{Ratio $R={4 \pi \Gamma_{K} \over |\Omega_{\bm{q} K}|}$ calculated at T = 50K for the three acoustic phonon bands: ZA, TA, LA for $L=68.4$nm.}
\label{R_Omegaqks_50K}
\end{centering}
\end{figure}

For $L={2 \pi \over q}=68.4$nm, both $T=50$K and $T=300K$ exhibit modes with $R_{\bm{q}K} \lesssim 1$ as shown in Figs. \ref{R_Omegaqks_300K}-\ref{R_Omegaqks_50K}. At $T=300$K, essentially only second phonons from the ZA branch with $\Omega_{\bm{q}K} < 0.15$THz propagate coherently, consistent with the observed spectra for $K^{\prime \prime}_{\bm{q}K}(\omega)$ shown in Fig. \ref{300K_imag}.
 Only in some limited cases are there values $R_{\bm{q}K} \sim 1$ in the LA and TA branches. At $T=50K$, Fig. \ref{R_Omegaqks_50K} demonstrates coherent second phonon propagation occurs in ZA, LA, and TA branches, although the ZA branch, with the smallest values of $R_{\bm{q}K}$, are expected dominate the second-sound response. The spectral results in Fig. \ref{50K_imag} and the breakdown by phonon branch in Fig. \ref{50K_imag_bands} shows this to be the case.

 In all cases, a significant number of modes have values $R_{\bm{q}K} \gg 1$, indicating that oscillations are strongly suppressed and the second phonon concept is not a useful picture. Recall that the frequency for second sound is determined from $\Omega_{\bm{q}K} = \bm{q} \cdot \bm{v}_{K}$, so that not only the magnitude  but also the direction of the $\bm{v}_{K}$ is important. Specifically, modes $K$ with $\bm{v}_{k}$ that have a large component perpendicular to $\bm{q}$ tend to correspond to low frequencies $\Omega_{\bm{q}K}$ and hence long periods. This results in a significant number of modes $K$ with $R_{\bm{q}K} \gg 1$ even in cases where $\Gamma_{K}$ is relatively small. For example, scattering rates $\Gamma_{K}$ for ZA modes can be quite low, yet many ZA modes exhibit $R_{\bm{q}K}$ values much larger than 1.

\begin{figure}
\begin{centering}
\includegraphics[scale=0.75]{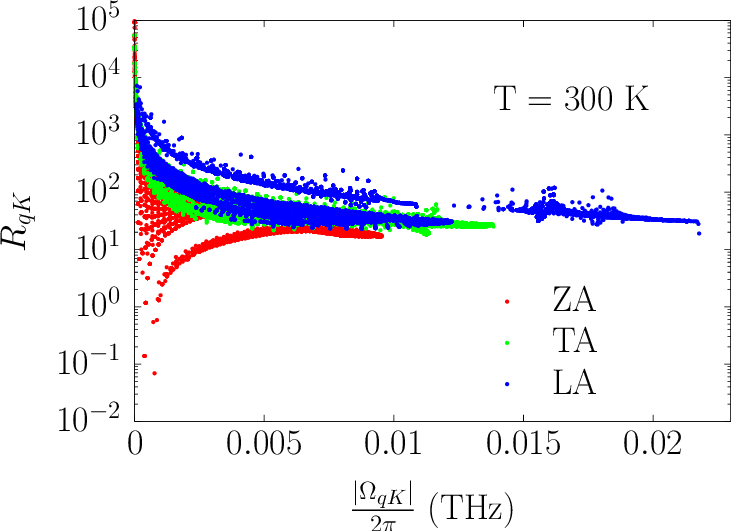} 
\caption{Ratio $R={4 \pi \Gamma_{K} \over |\Omega_{\bm{q} K}|}$ calculated at T = 300K for the three acoustic phonon bands. The length scale of the system is $L=1\mu$m.}
\label{R_Omegaqks_300K_1um}
\end{centering}
\end{figure}

\begin{figure}
\begin{centering}
\includegraphics[scale=0.75]{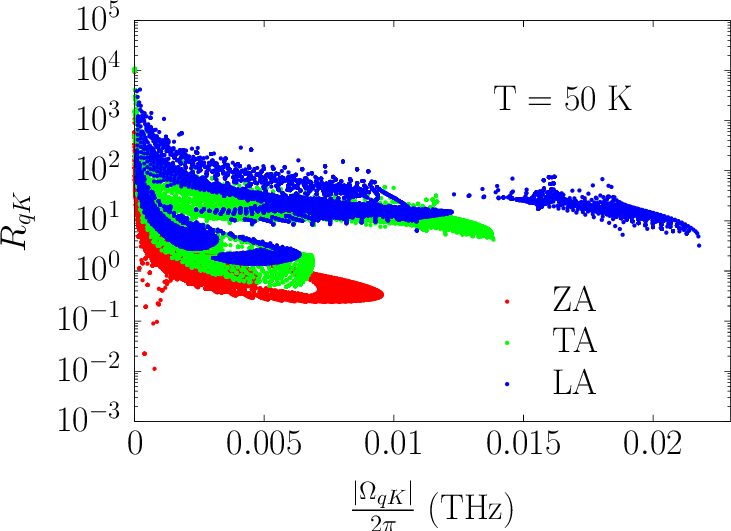} 
\caption{Ratio $R={4 \pi \Gamma_{K} \over |\Omega_{\bm{q} K}|}$ calculated at T = 50K for the three acoustic phonon bands: ZA, TA, LA for $L=1\mu$m.}
\label{R_Omegaqks_50K_1um}
\end{centering}
\end{figure}

The results for $R_{\bm{q}K}$ plotted for $L=1\mu$m in Fig. \ref{R_Omegaqks_300K_1um}-\ref{R_Omegaqks_50K_1um}. At $T=300$K, it is clear that only a very small portion of the ZA modes can propagate as second phonons. Hence, second sound at these scales is not observable as we have already demonstrated. At $T=50K$ and $L=1\mu$m, Fig. \ref{R_Omegaqks_50K_1um} shows that ZA modes should be able to propagate coherently with frequencies in the range $-0.01$THz to $0.01$THz, but potentially TA modes may also contribute in somewhat the same frequency range. However, as has already been demonstrated in Fig. \ref{50K_imag_bands_1um}, only the ZA portion of the spectrum exhibits coherent propagation. For $L=1\mu$m and $T=50$K, the TA and LA modes display only a limited number of modes with $R_{\bm{q}K} \sim 1$, which is not sufficient to propagate second sound.
  
  \begin{figure}
\begin{centering}
\includegraphics[scale=0.75]{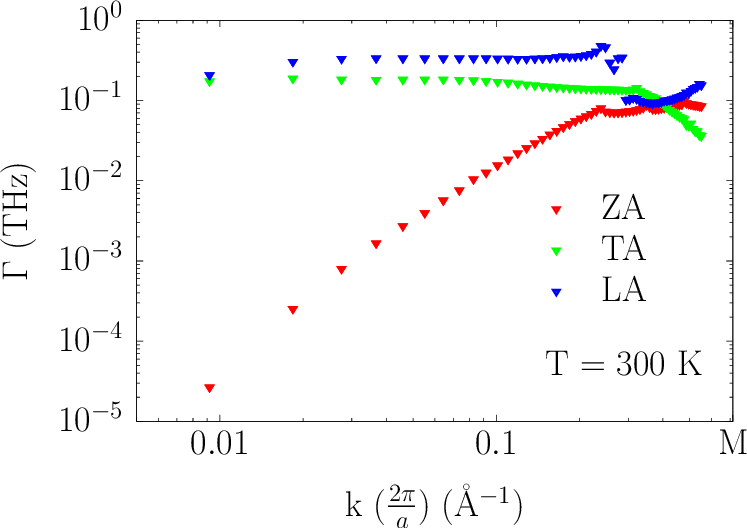} 
\caption{$\Gamma_{K}$ as a function of $k$ calculated at T = 300K for the three acoustic phonon bands. The unit is given in THz}
\label{gamma_k}
\end{centering}
\end{figure}

Finally, in Fig. \ref{gamma_k}, DFT values for the scattering rate $\Gamma_{K}$ at $T=300K$ are presented. At the low-$k$ long wavelength limit, the scattering rate for the ZA mode is about $2 \times 10^{-5}$THz. The LA and TA mode scattering rates are around $0.2$THz. These results are generally consistent with past results \cite{PhysRevB.89.155426,PhysRevB.108.L121412}. Using the relationship $\tau_{K}={1 \over 2 \Gamma_{K}}$ for the phonon lifetime, we obtain lifetimes $\sim 2.5$ps for LA phonons at long wavelengths. In comparison, scattering rates for TA modes are consistently lower. For ZA modes, lifetimes extend to $\sim 25$ns for the data shown in the figure, and depend strongly on $k$. Hence, the ZA branch is most able to sustain second phonon oscillations of the kind described here. As noted by other authors \cite{PhysRevB.89.155426,Bonini_2012,PhysRevB.108.L121412}, for the same reasons, ZA modes tend to dominate thermal conductivity as well.

  \section{Discussion and Conclusions}
  
  In this paper we have demonstrated that the common assumption in PBE theories of phonon hydrodynamics that the temperature field $T(\bm{r},t)$ and displacement field $\bm{V}(\bm{r},t)$ oscillate with a single frequency across the entire phonon spectrum likely breaks down especially for excitations with shorter periodicity (e.g. $L \lesssim 1\mu$m). By connecting a microscopic picture of heat wave propagation to a definition of the phase-space distribution function $f_{K}(\bm{r},t)$, the connection between solutions of the PBE and the concept of second phonons as elementary excitations is achieved. Solving the PBE separately for each mode results in a characteristic oscillation frequency $\Omega_{\bm{q}K}$ for each mode $K$. The spectral width of oscillations depends on the second-sound wave vector $\bm{q}$. Even for very weak scattering, the spectral width of the response function is limited by phonon dispersion. This is simply the physics of wave-packet spreading which was demonstrated in our earlier MD simulations \cite{Bohm:2022aa,Schelling:2025aa,Margolles_2026}, and altogether provides strong support for treating second sound as an elementary excitation along the lines of the ``second phonon'' envisioned by Sham \cite{Sham:1967aa,Sham:1967ab}. Finally, this approach shows how PBE theory can be amended to account for behavior at shorter length scales when interference effects due to the finite bandwidth of excitations $\Omega_{\bm{q}K}$ becomes important.

 One insight from our theory is to establish the dominant role of ZA phonons in second sound propagation. The spectral response functions primarily exhibit sharp, resolved features in the ZA modes. By contrast, LA and TA modes, characterized by higher scattering rates $\Gamma_{K}$, are responsible for broad, diffuse spectral features that are less likely to be observed as second sound. Consistent with our previous MD simulations \cite{Bohm:2022aa,Schelling:2025aa,Margolles_2026},  when scattering rates are low enough to lead to sharp spectral features, we find that the primary limitation in observing second sound is related to the bandwidth of the frequencies $\Omega_{\bm{q}K}$ for ZA phonons. This may also suggest that second sound is more easily observed in systems with a wide range of scattering rates $\Gamma_{K}$, since this should limit second phonons to a limited portion of the spectra.
  
However, our claim here is not that existing PBE theory fails to explain experiments. Specifically, despite not accounting for some of the physics elucidated here, existing theories based on the PBE nevertheless are successful at describing a wide range of experiments. We believe that this is due to the fact that most experiments so far have been performed for small values $\bm{q}$ corresponding to periodicity $L$ in the micron range. We note that for $L=1\mu$m, the computed response function for $T=50K$ in Fig. \ref{50K_imag_bands_1um} demonstrates a spectral width of only $\pm 0.01$THz. Therefore, the standard assumptions made in PBE theory that the displacement and temperature fields fluctuate with a single frequency is a reasonable approximation in this instance. Moreover, at these scales, the broadening of the response due to scattering can quickly overwhelm other factors. Consequently, it is likely that the predictions made here will become relevant primarily in experiments on shorter length scales, which can be realized using UV lasers in TTG experiments.

Recently, we have developed a theory based on Green's functions in Ref. \cite{Schelling:2025aa}. This is very close to the original work by Sham \cite{Sham:1967aa,Sham:1967ab}. This theory included vertex corrections which lead to coupling between Green's functions representing the  propagation of the second phonons. Importantly, the frequency response in that paper \cite{Schelling:2025aa} with the scattering matrix set to zero can be shown to be essentially equivalent to the work reported here in the RTA. Hence, the vertex corrections examined in Ref. \cite{Schelling:2025aa} represent an approach to extend the PBE theory of hydrodynamics and second sound beyond the RTA. This is the subject of future work. However, the encouraging point is that there appears to be a direct link between the PBE theory and the approach first suggested by Sham \cite{Sham:1967aa,Sham:1967ab}.

 \section{Appendix A}

 Here we make a connection between the microscopic energy density and the description based on a local phase density deviation $g_{K}(\bm{r},t)$ used in the PBE theory. This picture is developed entirely in the harmonic approximation without reference to scattering and exponential decays of the distribution $g_{K}(\bm{r},t)$.

The local energy is defined using a harmonic Hamiltonian, with momenta $p_{il}$ and displacements $u_{il\alpha}$,
\begin{equation}
H_{0} = \sum_{il} {p_{il}^{2} \over 2m_{i}}
+{1 \over 2} \sum_{il\alpha,jl^{\prime}\beta}
\phi^{(2)}_{il\alpha,jl^{\prime}\beta} u_{il\alpha}u_{jl^{\prime}\beta} \text{   ,}
\end{equation}
in which $i,j$ label atoms,  $l$ and $l^{\prime}$ label primitive unit cells, and $\alpha$, $\beta$ are labels for components in a Cartesian system. The term $\phi^{(2)}_{il\alpha,jl^{\prime}\beta}$ represents the harmonic force-constant matrix.
First we define the local energy term at site $il$. This is just,
\begin{equation}
E_{il} = {p_{il}^{2} \over 2m_{i}}
+ {1 \over 2} \sum_{\alpha,jl^{\prime}\beta}
\phi^{(2)}_{il\alpha,jl^{\prime}\beta} u_{il\alpha}u_{jl^{\prime}\beta}
\end{equation}
We are interested in the Fourier components of the energy density defined by the following. First for each primitive cell $l$, we define the energy density $U_{l}$
\begin{equation}
U_{l} = {1 \over v_{c}} \sum_{i} E_{il} = \sum_{\bm{q}} U_{\bm{q}} e^{i \bm{q} \cdot \bm{R}_{l}}
\end{equation}
in which $v_{c}$ is the volume of a primitive cell. Then the Fourier coefficients are found from,
\begin{equation}
U_{\bm{q}} = {1 \over N_{c}}\sum_{l} U_{l}e^{-i \bm{q} \cdot \bm{R}_{l}}
\end{equation}
in which $N_{c}$ is the number of primitive cells that comprise the system. Written in terms of the local energies $E_{il}$,
\begin{equation}
U_{\bm{q}} = {1 \over V} \sum_{il} E_{il} e^{-i \bm{q} \cdot \bm{R}_{l}}
\end{equation}
and also,
\begin{equation}
U_{-\bm{q}} = {1 \over V} \sum_{il} E_{il} e^{i \bm{q} \cdot \bm{R}_{l}}
\end{equation}
where $V= N_{c}v_{c}$ is the total system volume.

The momenta and displacements are written, taking $A_{\bm{k}s}$ to be real quantities and introducing a phase factor $e^{i \phi_{\bm{k}s}}$,
\begin{equation}
u_{il\alpha}(t)= \sum_{\bm{k}s}
\left( {\hbar \over 2m_{i} \omega_{\bm{k}s}}\right)^{1 \over 2} A_{\bm{k}s}\left[
 \epsilon_{\bm{k}s,i \alpha} 
e^{i\left( \bm{k} \cdot \bm{R}_{l} -\omega_{\bm{k} s} t\right)} e^{i \phi_{\bm{k}s}} +
 \epsilon_{-\bm{k}s,i \alpha} 
e^{-i\left( \bm{k} \cdot \bm{R}_{l} -\omega_{\bm{k} s} t\right)} e^{-i \phi_{\bm{k}s}}
\right]
\end{equation}
\begin{equation}
p_{il\alpha}(t) = -i \sum_{\bm{k}s}
\left( {m_{i} \hbar  \omega_{\bm{k}s}} \over 2  \right)^{1 \over 2} A_{\bm{k}s} \left[
\epsilon_{\bm{k}s,i \alpha} 
e^{i\left( \bm{k} \cdot \bm{R}_{l} -\omega_{\bm{k} s} t\right)} e^{i \phi_{\bm{k}s}} -
 \epsilon_{-\bm{k}s,i \alpha} 
e^{-i\left( \bm{k} \cdot \bm{R}_{l} -\omega_{\bm{k} s} t\right)} e^{-i \phi_{\bm{k}s}}
\right]  \text{   ,}
\end{equation}
where $\epsilon_{\bm{k}s,i \alpha}$ represents a component of the polarization vector for a 
phonon mode $(\bm{k}s)$.
We also use that $\epsilon_{-\bm{k}s,i \alpha} =\epsilon^{*}_{\bm{k}s,i \alpha} $.

With the definition of local energy, we can compute the energy that resides in unit cell $l$ from,
\begin{equation}
U_{l}(t) = \sum_{\bm{q}} \sum_{\bm{k}s} \hbar \omega_{\bm{k}s} A_{\bm{k} + {1\over 2} \bm{q}}A_{\bm{k} - {1\over 2} \bm{q}}
\left( \sum_{i \alpha} \epsilon_{\bm{k}+{1 \over 2} \bm{q}s;i\alpha} 
\epsilon_{-\bm{k}+{1 \over 2} \bm{q}s;i\alpha}\right)
  e^{-i \left( \omega_{\bm{k}+{1 \over 2} \bm{q}s} - \omega_{\bm{k} - {1 \over 2} \bm{q}s} \right)t}
    e^{i \left( \phi_{\bm{k}+{1 \over 2} \bm{q}s} - \phi_{\bm{k} - {1 \over 2} \bm{q}s} \right)} e^{i \bm{q} \cdot \bm{R}_{l}}
\end{equation}
Then the energy density in Fourier space is,
\begin{equation}
 U_{\bm{q}}(t) = {1 \over V} \sum_{\bm{k}s} \hbar \omega_{\bm{k}s}
A_{\bm{k} - {1 \over 2} \bm{q}s} A_{\bm{k}+{1 \over 2} \bm{q}s} 
  e^{-i \left( \omega_{\bm{k}+{1 \over 2} \bm{q}s} - \omega_{\bm{k} - {1 \over 2} \bm{q}s} \right)t}
    e^{i \left( \phi_{\bm{k}+{1 \over 2} \bm{q}s} - \phi_{\bm{k} - {1 \over 2} \bm{q}s} \right)}
 \end{equation}
 \begin{equation}
 U^{*}_{\bm{q}}(t) =U_{-\bm{q}}(t)  = {1 \over V} \sum_{\bm{k}s} \hbar \omega_{\bm{k}s}
  A_{\bm{k} + {1 \over 2} \bm{q}s} A_{\bm{k}-{1 \over 2} \bm{q}s} 
  e^{-i \left( \omega_{\bm{k}-{1 \over 2} \bm{q}s} - \omega_{\bm{k} + {1 \over 2} \bm{q}s} \right)t}
    e^{i \left( \phi_{\bm{k}-{1 \over 2} \bm{q}s} - \phi_{\bm{k} + {1 \over 2} \bm{q}s} \right)}
 \end{equation}
 The volume is the number of unit cells $N_{c}$ multiplying the volume of a cell, hence $V= N_{c}v_{c}$.
 Next let's write down $E_{l}(t)$, assuming only one $\bm{q}$ value is in resonance. Assuming that
 at $t=0$, all of the phases cancel (the differences above are zero). The deviation from the average density $U_{0}$ in unit cell $l$ is then,
  \begin{equation}
 U_{l}(t) -U_{0}={1 \over 2} \sum_{\bm{q}}\left[ U_{\bm{q}}(t) e^{i\bm{q} \cdot \bm{R}_{l}} +U_{-\bm{q}}(t) e^{-i\bm{q} \cdot \bm{R}_{l}}  \right]
 = {1 \over V} \sum_{\bm{q}} \sum_{\bm{k}s} \hbar \omega_{\bm{k}s}
  A_{\bm{k} + {1 \over 2} \bm{q}s} A_{\bm{k}-{1 \over 2} \bm{q}s} \cos{\left(\bm{q} \cdot \bm{R}_{l} - \Omega_{\bm{q},\bm{k}s}t 
  - \Delta \phi_{\bm{q},\bm{k}s} \right)}
 \end{equation}
 in which $\Omega_{\bm{q},\bm{k}s}=\omega_{\bm{k}+{1 \over 2} \bm{q}s} - \omega_{\bm{k}- {1 \over 2} \bm{q}s}$
 is the frequency difference between neighboring states in the phonon branch $s$ and $\Delta \phi_{\bm{q},\bm{k}s} = \phi_{\bm{k}+{1 \over 2} \bm{q}s}-\phi_{\bm{k}-{1 \over 2} \bm{q}s}$ represents the phase difference between neighboring modes in branch $s$.
   Thus, we see that what emerges is a 
 superposition of plane waves with wave-vector $\bm{q}$ but with a range of frequencies $\Omega_{\bm{q},\bm{k}s}$. Because the frequencies here represent a difference, we can, for small $|\bm{q}|$, then see that this represents a superposition of plane waves each traveling with their respective group velocities $\bm{v}_{\bm{k}s}$, hence,
   \begin{equation}
 U_{l}(t) -U_{0}=
{1 \over V} \sum_{\bm{q}}\sum_{\bm{k}s} \hbar \omega_{\bm{k}s}
  A_{\bm{k} + {1 \over 2} \bm{q}s} A_{\bm{k}-{1 \over 2} \bm{q}s} \cos{\left [\bm{q} \cdot \left(\bm{R}_{l} - \bm{v}_{\bm{k}s}t \right)
  - \Delta \phi_{\bm{q},\bm{k}s} \right]}
 \end{equation}

 From this, applying the continuity equation, the heat current is,
    \begin{equation}
    \bm{J}_{l}(t) = {1 \over V} \sum_{\bm{q}} \sum_{\bm{k}s} \hbar \omega_{\bm{k}s}  \bm{v}_{\bm{k}s} A_{\bm{k} + {1 \over 2} \bm{q}s} A_{\bm{k}-{1 \over 2} \bm{q}s}
    \cos{\left [\bm{q} \cdot \left(\bm{R}_{l} - \bm{v}_{\bm{k}s}t \right)
  - \Delta \phi_{\bm{q},\bm{k}s} \right]}
   \end{equation}

   For a particular vector $\bm{q}$, the second phonon frequency is $\Omega_{\bm{q},\bm{k}s} = \bm{q} \cdot \bm{v}_{\bm{k}s}$. Next, assuming that the phonon amplitudes vary smoothly in reciprocal space and that $|\bm{q}|$ is small, then it is reasonable to take the ``global'' (as opposed to local) occupancy of a phonon mode as $N_{\bm{k}s} = A_{\bm{k} + {1 \over 2} \bm{q}s} A_{\bm{k}-{1 \over 2}\bm{q}s} $. Also, in equilibrium, the
   phase terms $\Delta \phi_{\bm{q},\bm{k}s}$ should add randomly. Given a periodic heat pulse, phase angles $\Delta \phi_{\bm{q},\bm{k}s}$ will correspond to constructive interference and coherent phonon transport will occur. Finally, the equations above correspond entirely to harmonic transport. In the solutions to the PBE in the RTA, the anharmonic processes are added by including an exponential decay term for the deviation function $g_{K}(\bm{r},t)$. Hence, from the above expression for the heat-current density, we identify the time evolution of the deviation function, including the scattering in the RTA, as
   \begin{equation}
    g_{K}(\bm{r},t) = N_{K}\cos{\left(\bm{q}\cdot \bm{r}-\Omega_{\bm{q}K}t \right)}e^{-2\Gamma_{K}t}   
   \end{equation}

 \newpage
 


\begin{thebibliography}{10}

\bibitem{Guyer:1966aa}
R.~A. Guyer.
\newblock Solution of the linearized phonon boltzmann equation.
\newblock {\em Physical Review}, 148(2):766--778, 1966.

\bibitem{Guyer:1966ab}
R.~A. Guyer and J.~A. Krumhansl.
\newblock Thermal conductivity, second sound, and phonon hydrodynamic phenomena
  in nonmetallic crystals.
\newblock {\em Physical Review}, 148(2):778--788, 1966.

\bibitem{Hardy:1970aa}
Robert~J. Hardy.
\newblock Phonon boltzmann equation and second sound in solids.
\newblock {\em Physical Review B}, 2(4):1193--1207, 1970.

\bibitem{Cepellotti_2015}
Andrea Cepellotti, Giorgia Fugallo, Lorenzo Paulatto, Michele Lazzeri,
  Francesco Mauri, and Nicola Marzari.
\newblock Phonon hydrodynamics in two-dimensional materials.
\newblock {\em Nature Communications}, 6(1):6400, 2015.

\bibitem{Cepellotti:2016wi}
Andrea Cepellotti and Nicola Marzari.
\newblock Thermal transport in crystals as a kinetic theory of relaxons.
\newblock {\em Physical Review X}, 6(4):041013, 2016.

\bibitem{Lee:2017aa}
Sangyeop Lee and Lucas Lindsay.
\newblock Hydrodynamic phonon drift and second sound in a (20,20) single-wall
  carbon nanotube.
\newblock {\em Physical Review B}, 95(18):184304, 2017.

\bibitem{PhysRevB.101.075303}
A.~Beardo, M.~G. Hennessy, L.~Sendra, J.~Camacho, T.~G. Myers, J.~Bafaluy, and
  F.~X. Alvarez.
\newblock Phonon hydrodynamics in frequency-domain thermoreflectance
  experiments.
\newblock {\em Phys. Rev. B}, 101:075303, Feb 2020.

\bibitem{Simoncelli:2020aa}
Michele Simoncelli.
\newblock Generalization of fourier's law into viscous heat equations.
\newblock {\em Physical Review X}, 10(1), 2020.

\bibitem{Beardo_2021.2}
Albert Beardo, Joshua~L. Knobloch, Lluc Sendra, Javier Bafaluy, Travis~D.
  Frazer, Weilun Chao, Jorge~N. Hernandez-Charpak, Henry~C. Kapteyn, Bego{\~n}a
  Abad, Margaret~M. Murnane, F.~Xavier Alvarez, and Juan Camacho.
\newblock A general and predictive understanding of thermal transport from 1d-
  and 2d-confined nanostructures: Theory and experiment.
\newblock {\em ACS Nano}, 15(8):13019--13030, 2021.

\bibitem{Sendra:2022aa}
Lluc Sendra.
\newblock Hydrodynamic heat transport in dielectric crystals in the collective
  limit and the drifting/driftless velocity conundrum.
\newblock {\em Physical Review B}, 106(15), 2022.

\bibitem{Kwok:1966aa}
Philip~C. Kwok and Paul.~C. Martin.
\newblock Unified approach to interacting phonon problems.
\newblock {\em Physical Review}, 142(2):495--504, 1966.

\bibitem{Sham:1967aa}
L.~J. Sham.
\newblock Equilibrium approach to second sound in solids.
\newblock {\em Physical Review}, 156(2):494--500, 1967.

\bibitem{Sham:1967ab}
L.~J. Sham.
\newblock Temperature propagation in anharmonic solids.
\newblock {\em Physical Review}, 163(2):401--407, 1967.

\bibitem{Bohm:2022aa}
Nathaniel Bohm and Patrick~K. Schelling.
\newblock Analysis of ballistic transport and resonance in the
  alpha-fermi-pasta-ulam-tsingou model.
\newblock {\em Phys. Rev. E}, 106:024212, 2022.

\bibitem{MartinezMargolles:2025}
Antonio Martinez~Margolles and Patrick~K. Schelling.
\newblock Thermal response functions and second sound in graphene.
\newblock {\em https://arxiv.org/abs/2512.13988}, 2025.

\bibitem{Margolles_2026}
Antonio~Martinez Margolles and Patrick~K. Schelling.
\newblock Thermal response functions and second sound in graphene.
\newblock {\em Journal of Applied Physics}, 139(18):184302, 05 2026.

\bibitem{Schelling:2025aa}
Patrick~K. Schelling.
\newblock Thermal response functions and second sound in single-layer hexagonal
  boron nitride.
\newblock {\em Physical Review B}, 112(2), 2025.

\bibitem{Fernando_2020}
Kevin~M. Fernando and Patrick~K. Schelling.
\newblock Non-local linear-response functions for thermal transport computed
  with equilibrium molecular-dynamics simulation.
\newblock {\em Journal of Applied Physics}, 128(21):215105, 2020.

\bibitem{Kresse_1994}
G~Kresse and J~Hafner.
\newblock Norm-conserving and ultrasoft pseudopotentials for first-row and
  transition elements.
\newblock {\em Journal of Physics: Condensed Matter}, 6(40):8245--8257, October
  1994.

\bibitem{KRESSE199615}
G.~Kresse and J.~Furthm{\"u}ller.
\newblock Efficiency of ab-initio total energy calculations for metals and
  semiconductors using a plane-wave basis set.
\newblock {\em Computational Materials Science}, 6(1):15--50, 1996.

\bibitem{Kresse:1999aa}
G.~Kresse.
\newblock From ultrasoft pseudopotentials to the projector augmented-wave
  method.
\newblock {\em Physical Review B}, 59(3):1758--1775, 1999.

\bibitem{PhysRevB.54.11169}
G.~Kresse and J.~Furthm\"uller.
\newblock Efficient iterative schemes for ab initio total-energy calculations
  using a plane-wave basis set.
\newblock {\em Phys. Rev. B}, 54:11169--11186, Oct 1996.

\bibitem{Togo_2023.2}
Atsushi Togo.
\newblock First-principles phonon calculations with phonopy and phono3py.
\newblock {\em Journal of the Physical Society of Japan}, 92(1), January 2023.

\bibitem{Togo_2023}
Atsushi Togo, Laurent Chaput, Terumasa Tadano, and Isao Tanaka.
\newblock Implementation strategies in phonopy and phono3py.
\newblock {\em Journal of Physics: Condensed Matter}, 35(35):353001, 2023.

\bibitem{PhysRevB.89.155426}
L.~Lindsay, Wu~Li, Jes\'us Carrete, Natalio Mingo, D.~A. Broido, and T.~L.
  Reinecke.
\newblock Phonon thermal transport in strained and unstrained graphene from
  first principles.
\newblock {\em Phys. Rev. B}, 89:155426, Apr 2014.

\bibitem{PhysRevB.108.L121412}
Zherui Han and Xiulin Ruan.
\newblock Thermal conductivity of monolayer graphene: Convergent and lower than
  diamond.
\newblock {\em Phys. Rev. B}, 108:L121412, Sep 2023.

\bibitem{Bonini_2012}
Nicola Bonini, Jivtesh Garg, and Nicola Marzari.
\newblock Acoustic phonon lifetimes and thermal transport in free-standing and
  strained graphene.
\newblock {\em Nano Letters}, 12(6):2673--2678, may 2012.

\end{thebibliography}
\end{document}